\documentclass[12pt]{article}
\usepackage[margin=1in]{geometry}
\usepackage{setspace}
\usepackage[latin1]{inputenc}
\usepackage{amsmath}
\usepackage{amsfonts}
\usepackage{amssymb}
\usepackage{soul}
\usepackage{bm}
\usepackage{multirow}
\usepackage{tikz}
\usepackage{enumitem}  
\usetikzlibrary{shapes.geometric, arrows}
\usepackage[latin1]{inputenc}
\usepackage{filecontents}
\usepackage[small,bf]{caption}
\usepackage{graphicx,hyperref}
\usepackage[resetfonts]{cmap}
\hypersetup{colorlinks = true,linkcolor = blue,anchorcolor = blue,citecolor = blue,filecolor = blue,urlcolor = blue}

\newcommand{\vect}[1]{\boldsymbol{#1}}
\usepackage{lscape}
\usepackage{float}

\begin{document}

\pagenumbering{roman}
\begin{titlepage}
	\begin{center}
		\Large \textbf{Conditional logistic individual-level models of spatial infectious disease dynamics}\\
  \vspace{0.7cm}
		Tahmina Akter$^{1,2}$, Rob Deardon$^{1,3}$\\
		Department of Mathematics and Statistics, 
		University of Calgary$^1$\\	
  Faculty of Institute of Statistical Research and Training, University of Dhaka $^2$\\
  Faculty of Veterinary Medicine, University of Calgary$^3$
	\end{center}
	
	\vspace{0.7cm}

\section*{Abstract:}
Here, we introduce a novel framework for modelling the spatiotemporal dynamics of disease spread known as conditional logistic individual-level models (CL-ILM's). This framework alleviates much of the computational burden associated with traditional spatiotemporal individual-level models for epidemics, and facilitates the use of standard software for fitting logistic models when analysing spatiotemporal disease patterns. The models can be fitted in either a frequentist or Bayesian framework. Here, we apply the new spatial CL-ILM to both simulated and semi-real data from the UK 2001 foot-and-mouth disease epidemic.


\vspace{0.3cm}
\textit{Keywords:} Disease transmission model, ILMs, Logistic ILM, Conditional logistic ILM, Posterior predictive distribution.


\end{titlepage}

\pagenumbering{arabic}
\newpage
\section{Introduction}
Infectious disease outbreaks can have devastating effects on human lives, agriculture, and economic growth. For example, the ongoing coronavirus disease outbreak wreaked havoc on public health and economic activity lost (Barro et al., 2020). High-quality mathematical models can provide powerful insights into how infectious disease complex systems behave, which in turn can enable outbreaks to be better controlled by designing efficient public health strategies and resource allocation, such as intervention or vaccination (Tildesley et al., 2006). To this end, Deardon et al. (2010) introduced a class of individual-level models that focus on describing and predicting the behavior of disease at the individual level of interest (e.g., infection between people, households, or farms). 

Individual-level models are notable because they incorporate individual-specific covariate information on susceptible and infectious individuals to better describe the dynamics of infectious disease outbreaks. For example, we can account for population heterogeneity in space by including information on separation distance. However, fitting such models to data can be difficult due to the computational cost of calculating the likelihood. This situation arises when we deal with 
a large population. Utilizing ILMs is also challenging because it generally requires specialized software such as the EpiILM and EpiILMCT R packages (Warriyar et al., 2020; Almutiry et al., 2020) or coding in fast languages such as Fortran, Julia, or C. 

Inference for such models is usually facilitated via Markov chain Monte Carlo (MCMC) within a Bayesian framework. This is a powerful tool because it can deal with high-dimensional and complex models and offers great flexibility in the choice of model. Bayesian MCMC is also powerful in that it provides a principled way for imputing missing data, as well as enabling the incorporation of prior knowledge, allowing multiple sources of data to be combined to improve parameter identifiability. However, in terms of practical outcomes, repeating the calculation of likelihood for an ILM as required by the MCMC method can be computationally very expensive, especially when dealing with large population sizes or complex models (Deardon, 2010).

A logistic regression model is a powerful tool in statistics used for modelling binary response variables and prediction. It is used to model the relationship between predictor variables and binary responses. It can be used to predict the probability of an event occurring, such as disease status (yes/no), based on the associated predictors. Moreover, the logistic model can used as a valuable tool in epidemiology for understanding the dynamics of disease transmission within a population (Jin et al., 2015).
There is also, of course, a wide range of statistical software for fitting these models. The key features of these models are simplicity, interpretability, and applicability to a wide range of scenarios.

In this study, we propose a framework for logistic ILMs, specifically in the context of spatial individual-level models. The logistic ILM is used to model the probability of infection (or non-infection) of disease at each point in time based on risk factors (e.g., environmental, demographical, or behavioral) that are associated with individuals in the population. This is done in a similar way to an ILM, but the two models have a different underlying functional form.
 From these models, we can understand the spatial pattern of the disease, identify associated risk factors, and make predictions or forecasts just as we can with a standard ILM.
 
Spatial logistic ILMs are typically non-linear in terms of their covariates due to the spatial distance function typically used. However, we can condition on the spatial parameter of the logistic ILM so that the covariates in the model are linear predictors of the log odds of infection at each time point. This enables us to use standard statistical software 
to fit the logistic ILM and facilitates faster inference. 
We will do this in two stages. In the first stage, we will fix the spatial parameter by choosing an appropriate value for the parameter from a finite set of plausible values. This leads to a \textit{conditional logistic ILM (CL-ILM)}. In the second stage, the model can be fitted in either a Bayesian or frequentist framework. Here, we will focus on Bayesian CL-ILMs. We can check the performance of this model relative to say, a standard ILM by using a posterior predictive approach (Gardner et al., 2011), or model-based information criterion.

The subsequent sections of this paper are organized as follows. In Section 2, we introduce the general framework of ILMs, the logistic ILM, the spatial logistic ILM, the CL-ILM, methods for converting from epidemic data to that suitable for fitting the CL-ILM via standard statistical software, and the posterior predictive approach. In Section 3, we discuss our simulation process. In Section 4, we present our findings and compare the ILM and CL-ILM methods based on simulation studies under SI and SIR frameworks. In Section 5, we apply the CL-ILM to semi-real data based on the UK foot and mouth disease (FMD) outbreak of 2001. Finally, in Section 6, we conclude and propose plans for future research.

\section{Methodology}
\subsection{Individual-level model}
A class of disease transmission models defined as individual-level models was introduced by Deardon et al. (2010). These models provide a tool for modelling infectious disease spread through space and time at the individual level (e.g., individual people, households, or geographical regions). The goal of these models is to mimic the dynamic of infectious disease. The models are placed within a so-called compartmental framework.
We begin by considering the SI - or susceptible (S), infectious (I) framework and then the SIR - or susceptible (S), infectious (I), removed (R) framework. These compartmental frameworks can easily be extended to SEIR or SEIRS, which allows for a latent period and/or reinfections.

In the SI framework, individuals are initially in the susceptible state (S), and when infection occurs, the individual becomes infectious immediately and moves to the infectious state (I). In the SIR framework, 
the same process occurs but after some time the individual moves to the removal state (R), if death or recovery happens, for example. 
In our discrete-time scenarios, the epidemic starts at the time $t=1$ when the first individual is being infected and the epidemic ends at the time $t=t_{end}$; $t=1, 2, \ldots, t_{end}$. The functional form of the ILM infection probability as defined in Deardon et al. (2010) is given as,
\begin{align}
		P_{it}&= 1-\exp \left[-\big\{\Omega_S(i)\sum_{j\epsilon I(t)}\Omega_T(j) k(i,j)\big\}-\varepsilon (i,t)\right], \hspace{0.4cm} \Omega_S(i), \Omega_T(j), \varepsilon (i,t)>0
		\end{align}  
where: $P_{it}$ is the probability that susceptible individual $i$ is infected at time $t$; $I(t)$ is the set of individuals who are  infectious at time $t$; $\Omega_S(i)$ is a susceptibility function representing potential risk factors associated with the $i^{th}$ susceptible individual contracting the disease; $\Omega_T(j)$ is a transmissibility function representing potential risk factors associated with the $j^{th}$ infectious individual passing on the disease;
	$k(i,j)$ is an infection kernel that involves potential risk factors associated with both the infectious and susceptible individuals (e.g., a function of spatial distance); and $\varepsilon (i,t)$ describes random behavior due to some otherwise unexplained infection process. 
 
The likelihood function for the model of (1) is given as
	\begin{equation*}
		L(\vect{D}| \vect{\theta})= \prod_{t=1}^{t_{max}-1} f_t( \vect{D}| \vect{\theta}),
	\end{equation*}
	where
	\begin{equation*}
		f_t( \vect{D}| \vect{\theta})=\left[\prod_{i\epsilon I(t+1)\setminus I(t)}P_{it}\right] \left[\prod_{i\epsilon S(t+1)}(1-P_{it})\right],
	\end{equation*}
and where $\vect{\theta}$ is the vector of unknown parameters, $\vect{D}$ is the epidemic data set, $S(t+1)$ is the set of individuals susceptible at time $(t+1)$, $I(t+1)\setminus I(t)$ is the set of individuals newly infected at time $t+1$, and $t_{max} \le t_{end}$ is the last time point observed in data. Infectious periods (removal) can be modelled in various ways. For simplicity, here, we assume that the infection times and infection periods are known.  

We will focus upon a simple spatial ILM with no covariates aside from spatial distance with the form,
 \begin{align}
		P_{it}&=1-\exp \left[-\alpha\sum_{j\epsilon I(t)}d_{ij}^{-\beta}\right], \hspace{0.2cm}  \alpha, \beta>0, t=1, \ldots, t_{max},
		\end{align}  
where $\Omega_S(i)=\alpha$, $\Omega_T(j)=1$, $k(i,j)=d_{ij}^{-\beta}$ and $\varepsilon (i,t)=0$ in equation (2), and where $d_{ij}$ is the Euclidean distance between $i^{th}$ susceptible and $j^{th}$ infectious individual, $\alpha$ is the baseline susceptibility, and $\beta$ is the spatial parameter.

  \subsection{Logistic ILM}
  Here, we discuss the logistic ILM and its general form. The logistic ILM is the logistic version of the ILM that involves the relationship between log odds of infection and potential risk factors associated with susceptibility and transmissibility. The general form of the logistic ILM infection probability is defined as
  \begin{align}
      \lambda_{it}=\Psi_S \sum_{j\epsilon I(t)} \Psi_T K(i, j)+e_{it},
  \end{align}
  where $\lambda_{it}=\log \left[\frac{P_{it}}{1-P_{it}}\right]$, $\Psi_S$ is the potential risk factors associated with the susceptible individual, $\Psi_T$ is the potential risk factors associated with the infectious individual, $K(i, j)$ is the infection kernel, and $e_{it}$ is some infection from unexplained causes.
The likelihood function for the model (3) can be written as,
 \begin{align*}
     L(\vect{D}|\vect{\theta})=\prod_{i=1}^{n} \prod_{t=1}^{t_{max}-1} P_{it}^{y_{it}} (1-P_{it})^{1-y_{it}}, 
 \end{align*}
where $y_{it}$ is the infection status of $i^{th}$ individual at time $t$, with $y_{it}=1$ if the individual is infected and $0$ otherwise. We can fit this model to data by maximizing the likelihood and then take a frequentist approach to inference or fit the model in a Bayesian framework using an MCMC algorithm, incorporating prior information on parameters.

  \subsection{Spatial logistic ILM}
Here, we present a logistic version of the simple spatial ILM of equation (2). It can be considered as an alternative model in its own right, or as an approximation to the spatial ILM. 
It is given by,
\begin{align}
		\log \left[\frac{P_{it}}{1-P_{it}}\right]&=\vect{X}\vect{\alpha}
  =\alpha_0+\alpha_1 X_{it}
\end{align}  
where $\Psi_S=\vect{\alpha}$, $\Psi_T=1$, $K(i, j)=d_{ij}^{-\beta_0}$, and $e_{it}=0$ in equation (3), where
$\vect{X}=(1, X_{it})$, $\vect{\alpha^T}=(\alpha_0, \alpha_1)$ and $X_{it}=\sum_{j\epsilon I(t)}d_{ij}^{-\beta_0}$. That is we relate the force of infection $(\alpha_1\sum_{j\epsilon I(t)}d_{ij}^{-\beta_0})$ to the log odds of infection rather than the probability of infection.
We can write the probability of being infected as
 \begin{align*}
		P_{it}&=\frac{\exp(\vect{X}\vect{\alpha})}{1+\exp(\vect{X}\vect{\alpha})}\\
  &=\frac{\exp\big(\alpha_0+\alpha_1 X_{it}\big)} {1+\exp\big(\alpha_0+\alpha_1 X_{it}\big)}\\
  &=\frac{1}{1+\exp\Big(-\big(\alpha_0+\alpha_1 X_{it} \big)\Big)}.
		\end{align*}   

However, standard statistical software for fitting logistic models will not be able to cope with the non-linearity in the spatial function (e.g., the glm command in R), due to the non-linearity in $X_{it}$. However, if we fix, or condition on $\beta_0$, we can calculate $X_{it}$ for each susceptible $i$ at each time $t$ and then use standard software to fit the model.

\subsection{Conditional logistic ILM}
The conditional logistic ILM involves conditioning on the spatial parameter $\beta_0$. In such cases, the probability of being infected can be written as
 \begin{align*}		P(Y_{it}=1|\beta_0=\Tilde{\beta_0})&=\frac{\exp\big(\alpha_0+\alpha_1 \sum_{j\epsilon I(t)}d_{ij}^{-\Tilde{\beta_0}}\big)} {1+\exp\big(\alpha_0+\alpha_1 \sum_{j\epsilon I(t)}d_{ij}^{-\Tilde{\beta_0}}\big)}\\
  &=\frac{1}{1+\exp\Big(-\big(\alpha_0+\alpha_1 \sum_{j\epsilon I(t)}d_{ij}^{-\Tilde{\beta_0}} \big)\Big)},
		\end{align*} 
where $\Tilde{\beta_0}$ is our fixed value of $\beta_0$.
 The conditional likelihood function can be written as
 \begin{align}
     L(\vect{\alpha}|\beta_0)=\prod_{i=1}^{n} \prod_{t=1}^{t_{max}-1} P(Y_{it}|\Tilde{\beta_0})^{y_{it}} (1-P(Y_{it}|\Tilde{\beta_0}))^{1-y_{it}}.
 \end{align}

One simple way to choose $\Tilde{\beta_0}$ is to fit the model for each of a finite set of possibilities and choose the $\Tilde{\beta_0}$ which maximizes the likelihood.

\subsection{Data converting from epidemic data to binary data}
 The epidemic data for a spatial SI ILM without covariates will contain information on infection times (and removal times if an SIR model is being fitted) of individuals with $(X, Y)$ coordinates. Here, we will explain the procedure of how to convert such epidemic data to binary data that we can fit the spatial logistic ILM to. The infection pattern over time is shown in Table 1, for a hypothetical `toy example' consisting of four individuals. Note that the column `Individual ID' is not strictly needed but is included here to aid illustration.

\begin{table}[!h]
\centering
		\caption {Infection pattern over time.}
\begin{tabular}{|c|c|c|}
\hline
 Individual ID & $(X,Y)$ coordinate & ``Infectious'' time \\
 \hline
1        & (2.6, 1.5)                             & 5       \\
2       & (3.7, 6.8)                              & 4          \\
3         & (5.7, 6.5)                            & 2         \\
4           & (5.9, 6.3)                          & 3    \\
\hline    
\end{tabular}
\end{table}
In this data, individual 1 is being infected at $t=4$, and so becomes infectious at $t=5$, and similarly, individual 2 is being infected at $t=3$, and so becomes infectious at $t=4$, and so on. Here, the epidemic starts when individual 3 becomes infectious at time $t=2$, and we would condition on that infection. 

To convert the epidemic data to a data set suitable for fitting the CL-ILM using standard software, we create three columns. The first column incorporates time points for each individual. The time point will reach up to the point they get infected. The second column incorporates the infection event status of the individuals for each time point. We start to observe the binary data from $t=2$ because the epidemic starts from one individual who was infected at time $1$. The third column includes the set of infected individuals $(I_t)$ in the data. It contains $X_{it}$ calculated for fixed $\Tilde{\beta_0}$. Note that, $X_{it}$ will typically be calculated for change over time for each individual. The binary data set for Table 1 is shown in Table 2. Here, time $(t)$ and $I_t$ are supporting information that is not directly used in the fitting of our CL-ILM.

\begin{table}[!h]
\centering 
		\caption {The binary epidemic data. }
\begin{tabular}{|c|c|c|c|c|c|}
\hline
 Individual ID & (X,Y) coordinate & Time $(t)$  & Infection & $I_t$ & $X_{it}$\\
 \hline
1        & (2.6, 1.5)                              & 2 & 0 & \{3\}  &  $\sum_{j\epsilon I(2)}d_{1j}^{-\Tilde{\beta_0}} $    \\
1      & (2.6, 1.5)                                & 3 & 0 & \{3,4\}     &  $\sum_{j\epsilon I(3)}d_{1j}^{-\Tilde{\beta_0}}$    \\
1     & (2.6, 1.5)                                 & 4 & 1 & \{2,3,4\}     &  $\sum_{j\epsilon I(4)}d_{1j}^{-\Tilde{\beta_0}} $    \\
2      & (3.7, 6.8)                                  & 2&0&\{3\}     &  $\sum_{j\epsilon I(2)}d_{2j}^{-\Tilde{\beta_0}} $     \\
2      & (3.7, 6.8)                                  & 3&1&\{3,4\}  &  $\sum_{j\epsilon I(3)}d_{2j}^{-\Tilde{\beta_0}}  $       \\
4     & (5.9, 6.3)                                 & 2&1&\{3\}   &  $\sum_{j\epsilon I(2)}d_{4j}^{-\Tilde{\beta_0}} $ \\
\hline    
\end{tabular}
\end{table}

Similarly, we can convert the epidemic data to binary data in the context of the SIR framework. In this case, the data contains the information on the time of infection and time of removal with $(X, Y)$ coordinates for each individual. An infectious individual would move to the removal state after their infectious period. At that time, the individual would not be in the set of infectious individuals anymore, and this would feature in the calculation of $X_{it}$.

 \subsection{Posterior predictive distribution}
 To investigate the model accuracy or goodness of fit under the Bayesian framework, we can use a posterior predictive approach as introduced by Guttman (1967). 
 We can generate realizations from the posterior predictive distribution (PPD) of various epidemiological statistics such as some form of epidemic curve or the final size of the epidemic, and then compare that with the equivalent statistic calculated from the observed data, to assess the model fit. Here, we consider the number of newly infectious individuals (incidence) over time, which we refer to as the epidemic curve.
The algorithm for producing posterior predictive realizations in the case of an ILM or CL-ILM consists of the following steps:
\begin{enumerate}[label=(\roman*)]
    \item Sample a set of parameters from the MCMC-estimated posterior distribution.
    \item Simulate an epidemic from the model using the parameters sampled in Step (i).
    \item Summarize the simulated epidemic from Step (ii) via the epidemic curve (or some other statistic of interest).
    \item Repeat Steps (i) to (iii) a large number of times. For this study, we repeated 500 times.
 \end{enumerate}

Then, we examine and compare the PPD of the epidemic curve to the original observed epidemic curve to check for accuracy and precision. We consider a model to be a good fit for the data if the observed data lies in the areas of high mass of the PPDs, and the PPD has low variance.

To quantify the posterior predictive model fit, we can also use metrics such as the mean square error (MSE). Here, the MSE is calculated by taking the average of the squared differences between predicted and actual values of new infections over time, which is then averaged over the total number of epidemic simulations.
The MSE is given as,
\begin{equation*}
 MSE=\frac{1}{500t_{max}}\sum\limits_{s=1}^{500}\sum\limits_{t=1}^{t_{max}}(Y_{st}-\hat{Y}_{st})^2, 
\end{equation*}
where $Y_{st}$ is the number of new cases of the $s^{th}$ sample at time $t$, and $\hat{Y}_{st}$ is the predicted number of new cases of the $s^{th}$ sample at time $t$.

\section{Simulation study}
A simulation study is carried out to assess the performance of our CL-ILMs when the underlying data is generated by the spatial ILM (equation 2). That is, we examine how well a spatial CL-ILM can approximate the basic spatial ILM. Here, we consider a log transformation of the $X_{it}$ covariate to enhance stability in the data. Each data analysis is carried out in two stages. In the first stage, we use maximum likelihood over a finite set of $\beta_0$ values to tune $\beta_0$.
In the second stage, we fit the CL-ILM under the Bayesian framework. Then, we examine the model accuracy via the posterior predictive approach described above. Moreover, we compare the prediction error between the basic spatial ILM and CL-ILM under the SI and SIR frameworks. 

In this study, we simulate epidemic data under four scenarios with different spatial ILM parameter values. The true parameter values of $(\alpha, \beta)$ are $(0.7, 4)$, $(0.5, 3)$, $(0.2,4)$, and  $(0.9, 5)$ for each of the four scenarios, respectively. For each set of parameters, we produce 30 epidemics. These are used to fix $\beta_0=\Tilde{\beta_0}$. We then take an arbitrarily chosen subset of 20 epidemics and fit the CL-ILM to these using a Bayesian MCMC framework. Note, the subset of only 20 is taken at the second stage to reduce the computational burden associated with carrying out multiple MCMC analyses. For each simulated epidemic, we randomly
generate the spatial location of 500 individuals uniformly within $10 \times 10$ unit square area for each epidemic.
To generate epidemic data from the ILM, we use the epidata function from the `EpiILM' R package. Then, we convert the epidemic data to binary data suitable for analysing with the glm command in R.

 
\subsection{Fixing $\beta_0$}
We compare a number of spatial logistic ILMs to find the optimal tuning parameter ($\beta_0$). For comparing these models, maximizing the likelihood approach is used here. We compare the logistic models with spatial parameter $\beta_0\in \{-1, 0.5, 1, \ldots, 9.5, 10\}$. By fixing $\beta_0$, we construct the conditional logistic model. 


We fit models and calculate the likelihood values using the glm function in R with a logit link. We also calculate the proportion of time each possible value of $\beta_0$ is selected.

\subsection{Model fit}
Here, we fit the basic spatial ILM  using the mcmc function of the package `adaptMCMC' in R. For the basic spatial ILM, the marginal prior distributions of the parameters, $\alpha$ and $\beta$, are $U(0, 5)$ and $U(0, 10)$, respectively. Posterior predictive simulations are produced using the epidata command in the `EpiILM' R package.

In the case of CL-ILM, we fit the model using the MCMClogit function from the `MCMCpack' package in R. Here, the marginal prior distribution of the parameters, $\alpha_0$ and $\alpha_1$, are independent Cauchy distributions with location and scale parameters 0 and 1, respectively. 
We use our own R code to produce the epidemic curves under the posterior.
Then we compare the fit of the spatial ILM and the CL-ILMs. We use the average MSE to measure the prediction error and the average standard deviation (SD) to capture the variation in the posterior realizations. Moreover, we report the average proportion of time points at which posterior predictive 95\% credible intervals capture the true numbers of new infections.

\section{Results}
\subsection{SI framework}
Under the SI compartmental framework, we assess the performance of CL-ILMs when data is generated from a basic spatial ILM.

\subsubsection{Choosing $\beta_0$ via maximizing the likelihood}
For the true parameter $(\alpha, \beta)=(0.7, 4)$, the spatial parameter $\Tilde{\beta_0}$ was found to be either 3.5, 4.0, or 4.5 by maximizing the likelihood (Table 3). Further, $\Tilde{\beta_0}=4.0$ (the true value) was chosen under a majority of epidemics $(0.60)$. Similarly, when the true parameter value was $(0.5, 3)$, the highest proportion was 0.733 for the true value of $\Tilde{\beta_0}=3.0$. When the true values were $(0.2, 4)$ and $(0.9,5)$, the highest proportion was 0.533 and 0.600 for $\Tilde{\beta_0}=4.0$ and $\Tilde{\beta_0}=5.0$, respectively.

\begin{table}[!h]
\centering
		\caption {Proportion of time $\Tilde{\beta_0}$ selected for the four scenarios under the SI framework. }
\begin{tabular}{|c|c|c|}
\hline
  True parameter values $(\alpha, \beta)$ & Selected $\Tilde{\beta_0} $& Proportion  \\
 \hline
  (0.7, 4)              & 3.5                       & 0.133        \\
                        & 4.0                     & 0.600        \\
                        & 4.5                      & 0.267       \\
 \hline
  (0.5, 3)              & 3.0                     & 0.733        \\
                        & 3.5                     & 0.267        \\
 \hline
  (0.2, 4)              & 4.0                      & 0.533        \\
                        & 4.5                      & 0.467        \\
 \hline
  (0.9, 5)              & 4.5                        & 0.167        \\
                        & 5.0                       & 0.600        \\
                        & 5.5                      & 0.167        \\
                        & 6.0                      & 0.067       \\
 \hline
\end{tabular}
\end{table}

This implies that the maximizing likelihood approach can be successfully used to fix $\Tilde{\beta_0}$ for the CL-ILM, either picking the true or one close to the true value under all epidemic scenarios tested.
\newpage
\subsubsection{Model fit}
Table 4 shows the average MSE and SD under posterior prediction of the incidence-based epidemic curves under the SI framework. For each scenario, the average MSE and SD were higher for the CL-ILM compared to the spatial ILM. This would be expected, of course, since the actual data observed was simulated from the ILM.

\begin{table}[!h]
 \centering
		\caption {Comparing average MSE and average SD between the ILM and CL-ILM under the SI framework. }
\begin{tabular}{|c|c|c|c|c|}
\hline
True parameter values  & \multicolumn {2}{|c|}{ILM }             & \multicolumn {2}{|c|}{CL-ILM}         \\
\cline{2-3} 
\cline{4-5}
   $(\alpha, \beta)$    & Avg(MSE)      & Avg(SD)       & Avg(MSE)      & Avg(SD)  \\
 \hline
 
 (0.7,4)               & 596.626   & 601.064  & 840.756      & 629.226  \\
(0.5,3)               & 625.753   & 668.041  & 836.540      & 792.532 \\
(0.2, 4)              & 514.355  & 426.319 & 694.453     & 455.477  \\
(0.9, 5)              & 491.494   & 415.214  & 729.006      & 436.251 \\ 
 \hline
\end{tabular}
\end{table}

We summarize the mean proportion of time credible intervals capture the true number of infectious with its standard deviation across epidemic datasets in Table 5. When comparing the ILM and CL-ILM, we observed that the mean proportion value was slightly lower for the CL-ILM compared to the ILM. The mean proportion of successful incidence capture varied between 0.968 and 0.992 while considering the spatial ILM. In contrast, the mean proportion varied between 0.881 and 0.950 while considering the CL-ILM.
As expected, the values of SD were higher for the CL-ILM compared to the spatial ILM. However, we note that under the CL-ILM the lowest capture proportion was 0.881, and in the other scenarios it was larger than 0.90.

\begin{table}[!h]
 \centering
		\caption {Mean proportion of time credible interval captures the true distribution of infection with its SD under the SI framework. }
\begin{tabular}{|c|c|c|c|c|}
\hline
True parameter values  & \multicolumn {2}{|c|}{ILM }             & \multicolumn {2}{|c|}{CL-ILM}         \\
\cline{2-3} 
\cline{4-5}
  $(\alpha, \beta)$      & Mean & SD & Mean & SD \\
        \hline
(0.7, 4)               & 0.992           & 0.037          & 0.912           & 0.148           \\
(0.5, 3)               & 0.990            & 0.045           & 0.950            & 0.110           \\
(0.2, 4)              & 0.968           & 0.082           & 0.881            & 0.178          \\
(0.9, 5)              & 0.973            & 0.072           & 0.911            & 0.173      \\
 \hline
\end{tabular}
\end{table}

\newpage
From Figures 1 to 4 (see Appendix), we show the comparison of the posterior predictive epidemic curve (number of newly infectious individuals over time) between the ILM and CL-ILM for each scenario under the SI framework. 
We observed that the width of the posterior predictive intervals was a little larger under the ILM than the CL-ILM for each scenario, suggesting less uncertainty in epidemic prediction under the CL-ILM. This is presumably due to the fixing of the spatial parameter.
As we have already observed, there is also a higher chance of failing to capture the true incidence under the CL-ILM. However, the patterns of the posterior predictive distributions were fairly similar under the ILM and CL-ILM. Overall, this suggests that the CL-ILM provides a reasonable approximation to the basic spatial ILM.

\subsection{SIR framework}
Under the SIR compartmental framework, we evaluate the performance of CL-ILMs when data is generated from a basic spatial ILM. Here, we consider the infectious period follows Poisson distribution with a mean value of 4. 

\subsubsection{Choosing $\beta_0$ via maximizing the likelihood}
For the true parameter value of $(\alpha, \beta)=(0.7, 4)$, the spatial parameter $\Tilde{\beta_0}$ was found to be either $3.5$, $4.0$, $4.5$, or $5.5$ by maximizing the likelihood (Table 6). The highest proportion was 0.500 for the true value of $\Tilde{\beta_0}=4.0$. Similarly, when the true parameter value was $(0.5, 3)$, for the majority of epidemics ($0.667$) $\Tilde{\beta_0}=3.0$ was chosen. When the true values were $(0.2, 4)$ and $(0.9,5)$, the highest proportion was 0.633 and 0.400 for $\Tilde{\beta_0}=4.0$ and $\Tilde{\beta_0}=5.0$, respectively.

\begin{table}[!h]
\centering
		\caption {Proportion of time $\Tilde{\beta_0}$ selected for the four scenarios under the SIR framework. }
\begin{tabular}{|c|c|c|}
\hline
  True parameter values $(\alpha, \beta)$ & Selected $\Tilde{\beta_0}$ & Proportion  \\
 \hline
  (0.7, 4)              & 3.5                      & 0.133        \\
                        & 4.0                        & 0.500         \\
                        & 4.5                      & 0.333       \\
                        & 5.5                       & 0.033       \\
 \hline
  (0.5, 3)              & 2.5                       &  0.033        \\
                        & 3.0                    &  0.667       \\
                        & 3.5                     & 0.233        \\
                        & 4.0                      & 0.067        \\
 \hline
  (0.2, 4)              & 3.5                      & 0.033        \\
                        & 4.0                       & 0.633          \\
                        & 4.5                      & 0.300        \\
                        & 5.0                     & 0.033         \\
 \hline
  (0.9, 5)              & 4.5                       & 0.200        \\
                        & 5.0                       & 0.400        \\
                        & 5.5                      & 0.200       \\
                        & 6.0                      & 0.167       \\
                         & 6.5                      & 0.033     \\
 \hline
\end{tabular}
\end{table}

Once again, the findings imply that the maximizing likelihood approach can be effectively used to fix $\Tilde{\beta_0}$ for the CL-ILM, either picking the true value or one close to the true value under all epidemic scenarios tested.

\newpage
\subsubsection{Model fit}
Table 7 shows the average MSE and SD under posterior prediction of the incidence-based epidemic curves under the SIR framework.
Once again, the average MSE and SD were higher for the CL-ILM compared to the spatial ILM for all scenarios. This would be anticipated, of course, since the real data observed was simulated from the ILM.
\begin{table}[!h]
 \centering
		\caption {Comparing average MSE and average SD between the ILM and CL-ILM under the SIR framework. }
\begin{tabular}{|c|c|c|c|c|}
\hline
True parameter values  & \multicolumn {2}{|c|}{ILM }             & \multicolumn {2}{|c|}{CL-ILM}         \\
\cline{2-3} 
\cline{4-5}
   $(\alpha, \beta)$    & Avg(MSE)      & Avg(SD)       & Avg(MSE)      & Avg(SD)  \\
 \hline
 
 (0.7,4)               &716.932   & 644.098  & 1315.145     & 1105.329  \\
(0.5,3)               & 609.157   & 676.928  & 889.988      & 897.006 \\
(0.2, 4)              & 600.913   & 507.291  & 2811.673     & 1304.466 \\
(0.9, 5)              &  612.103   & 494.885  & 2472.659     & 1544.632 \\ 
 \hline
\end{tabular}
\end{table}

 In Table 8, we summarise the mean proportion of time points at which credible intervals capture the true number of infections with its standard deviation under the SIR framework. For each scenario, the mean proportion value was slightly lower for the CL-ILM compared to the spatial ILM. Under the CL-ILM, the lowest capture proportion was $0.843$, and the highest capture proportion was $0.950$. Under the ILM, the mean proportion varied between $0.959$ to $0.990$.
 Moreover, the standard deviations were higher for the CL-ILM compared to the ILM for all scenarios except $(0.9, 5)$. 

\begin{table}[!h]
 \centering
		\caption {Mean proportion of time credible interval captures the true distribution of infection with its SD under the SIR framework. }
\begin{tabular}{|c|c|c|c|c|}
\hline
True parameter values  & \multicolumn {2}{|c|}{ILM }             & \multicolumn {2}{|c|}{CL-ILM}         \\
\cline{2-3} 
\cline{4-5}
     $(\alpha, \beta)$   & Mean & SD & Mean & SD \\
        \hline
(0.7,4)               & 0.9590           & 0.0735          & 0.8924           & 0.1090          \\
(0.5,3)               & 0.9900           & 0.0447          & 0.9500           & 0.0784         \\
(0.2, 4)              & 0.9866           & 0.0413          & 0.8459           & 0.0698          \\
(0.9, 5)              & 0.9661           & 0.0605          & 0.8432           & 0.0418       \\
 \hline
\end{tabular}
\end{table}

Figures 5 to 8 (see Appendix) show the posterior predictive distribution of the epidemic curves under the ILM and CL-ILM for each scenario under the SIR framework. Here, we notice that the width of the posterior predictive intervals is larger for the CL-ILM compared to the ILM for almost all scenarios, suggesting more uncertainty in epidemic prediction under the CL-ILM.
Moreover, the patterns of the posterior predictive distributions are slightly different for the CL-ILM compared to ILM. Note that this differs from performance under the SI model.
However, the credible interval mostly captures the true number of infections, suggesting that the CL-ILM provides a reasonable approximation to the basic spatial ILM.

\section{Semi-real data}
Here, we fit the CL-ILM to a simulated epidemic based on foot and mouth disease data (FMD) from the UK  epidemic of 2001. The reason for using this `semi-real' data rather than the actual data set is that the culling strategy imposed by the UK government in 2001 is very hard to mimic, and so the posterior predictive performance of the epidemic model tends to be poor. The culling strategy varied over time and space but essentially aimed at pre-emptively culling animals as farms thought to be at high risk. Thus, we simulate a new `true' epidemic under our ILM that does not involve a culling strategy when fitted to this data. Then we compare the performance of the ILM and CL-ILM based on the `semi-real' data. We consider a subset of $1101$ farms from the Cumbria region, with infection times varied between $t=30$ to $71$ in days ($t=1$ being the day of the first infection).

In this study, we consider a conditional logistic ILM of the following form,
\begin{align*}
		\log \left[\frac{P(Y_{it}|\Tilde{\beta_0})}{1-P(Y_{it}|\Tilde{\beta_0})}\right]&=\alpha_0+\alpha_1 \sum_{j\epsilon I(t)}d_{ij}^{-\Tilde{\beta_0}},
		\end{align*} 
where: $\alpha_0$ is the intercept and $\alpha_1$ is the slope of the model.
Then we compare the model with the spatial ILM as follows
 \begin{align*}
		P_{it}&=1-\exp \left[-\alpha\sum_{j\epsilon I(t)}d_{ij}^{-\beta}\right] .
		\end{align*}  

To simulate the epidemics, we used true parameter values $(\alpha, \beta)=(0.00096, 1.22)$ and $(\alpha, \beta)=(0.002, 1.18)$ under the SI and SIR framework, respectively. The parameter values were estimated from the real FMD data using the optim function in R. In the case of SIR, we consider the infectious period following the Poisson distribution with a mean of 8.86. The mean value was the average infectious period found in the real FMD data. 

We compared the spatial logistic ILMs by maximizing likelihood values to fix $\beta_0$. Here, the values of $\beta_0$ considered were $\left\{-1, 0.2, 0.4, \ldots, 3.8, 4.0 \right\}$.
The spatial parameter $\Tilde{\beta_0}$ was found $1.0$ and $1.2$ under the SI and SIR framework, respectively.
This implies that the maximizing likelihood approach can be successfully used to fix $\Tilde{\beta_0}$ for the CL-ILM by picking a value close to the true value.

Then, we assess the performance of CL-ILM under the posterior and compare it with the basic spatial ILM.
Table 9 and Table 10 show the average MSE, and SD for posterior prediction of the incidence-based epidemic curves in the SI and SIR framework, respectively. In addition, we summarize the proportion of time credible intervals capture the true number of new infections in these tables.

\begin{table}[!h]
 \centering
		\caption {Average MSE, SD, and proportion of time original distribution capture in the credible intervals under the SI framework}
\begin{tabular}{|l|c|c|}
\hline
               & ILM    & CL-ILM \\
               \hline
Avg MSE        & 13.368 & 13.257      \\
SD         & 7.087 & 7.850        \\
Proportion & 1.000     & 1.000    \\
\hline
\end{tabular}
\end{table}

The average MSE and SD were very close for the spatial ILM and CL-ILM when considering the SI framework (Table 9). we observed that the proportion was exactly one for both the spatial ILM and CL-ILM. 
In contrast, the average MSE and SD were higher for the CL-ILM compared to the spatial ILM when considering the SIR framework (Table 10). The proportion of capturing original distribution was slightly higher for the spatial ILM ($1.000$) compared to the CL-ILM ($0.976$). Note that we simulated more epidemics under the semi-real data scenario and the findings were very similar.

\begin{table}[!h]
 \centering
		\caption {Average MSE, SD, and proportion of time original distribution capture in the credible intervals under the SIR framework}
\begin{tabular}{|l|c|c|}
\hline
               & ILM    & CL-ILM \\
               \hline
Avg MSE        & 46.716 & 83.621      \\
SD         & 35.259 & 42.533      \\
Proportion & 1.000  & 0.976 \\
\hline
\end{tabular}
\end{table}

Figure 9 (see Appendix) demonstrates the posterior predictive distribution and 95\% credible interval of the spatial ILM and CL-ILM for the semi-real data. The patterns of the posterior predictive distribution were fairly similar for both the spatial ILM and CL-ILM.
In addition, the posterior uncertainty was almost the same for the ILM and CL-ILM in the context of the SI framework. Alternatively, the posterior uncertainty was slightly higher for the CL-ILM compared to the ILM in the context of the SIR framework. Overall, this suggests that the CL-ILM is a reasonable approximation to the basic spatial ILM.
 

\section{Discussion}
 This article has proposed a logistic ILM as both an alternative to, and approximation of, the individual-level model. Generally, the ILM is a complicated model and thus the inference for these models is computationally expensive especially when involves a large population. Moreover, the ILM generally calls for coding in low-level language which makes the analysis harder for the researcher with limited expertise in computational statistics. We use a new modelling framework called CL-ILMs. The logistic model is a well-understood model with an extensive choice of statistical software for fitting into data, and the CL-ILM is associated with a substantially lower computational burden.

We use the posterior predictive approach to compare the performance of the CL-ILM when approximating a basic spatial ILM. To quantify prediction accuracy, we measured MSE and standard deviation. We discuss and compare the performance in the context of spatial disease models with simulated datasets and semi-real data from the UK 2001 foot-and-mouth disease epidemic. Overall, we find reasonably good prediction accuracy for the CL-ILM when comparing it with the spatial ILM. However, the posterior predictive uncertainty was found to be greater under the SIR framework compared to the SI framework. 

Of course, this study has some limitations and there are other avenues of research worthy of exploration. Firstly, we supposed that event times (infection and removal times) are known. However, the event times are typically not observed in practice with MCMC being used to solve this issue. However, it would be recommended to validate that our conclusions are robust when allowing for uncertain event times, though this would undoubtedly increase computation costs.
Secondly, we use the maximize likelihood approach for tuning the spatial parameter in the CL-ILM. However, other methods such as probability scoring rules could be considered here. In addition, if susceptibility and/or transmissibility covariates are being included in the model, then the choice of fixed spatial parameter will need to incorporate model uncertainty regarding the covariates. Thus, we might want to consider criteria such as AIC or BIC for a few covariates or methods such as the LASSO or spike-and-slab priors with large numbers of covariates.

Finally, here we have only considered SI and SIR compartmental frameworks for our CL-ILM, but extension to others, such as the SEIR would be warranted. We can also consider the introduction of more complex data structures and dynamics into our CL-ILM framework. For example, we could consider behaviour change mechanisms (e.g., Ward et al., 2023), population incorporating regional as well as individual-level spatial information (e.g., Mahsin et al., 2022), missing covariate information (Amiri et al., 2023) and contact network based continuous time ILMs (Almutiry \& Deardon, 2020).

\subsection*{CRediT authorship contribution statement}

\textbf{First Author (Corresponding Author):} Conceptualization, Formal analysis, Methodology, Software, Visualization, Writing - original draft, Writing - review \& editing.\\
\textbf{Second Author:} Conceptualization, Methodology, Supervision, Validation, Visualization, Writing - review \& editing.

\subsection*{Declaration of competing interest}
The authors declare that they have no known financial conflicts of interest or personal connections that might have influenced the work reported in this paper.

\subsection*{Acknowledgements}
This project was funded by an Alberta Innovates Graduate Student Scholarship for Data-Enabled Innovation and a University of Calgary Eyes High Doctoral Scholarship, Doctoral Completion Scholarship, Natural Sciences and Engineering Research Council of Canada (NSERC) Discovery Grants program (RGPIN/03292-2022) and the Alberta Innovates Advance - NSERC Alliance program (222302037).

\newpage
\section*{Appendix}
\subsection*{SI}
\begin{figure}[H]
		\centering
		\includegraphics[width=0.9\textwidth]{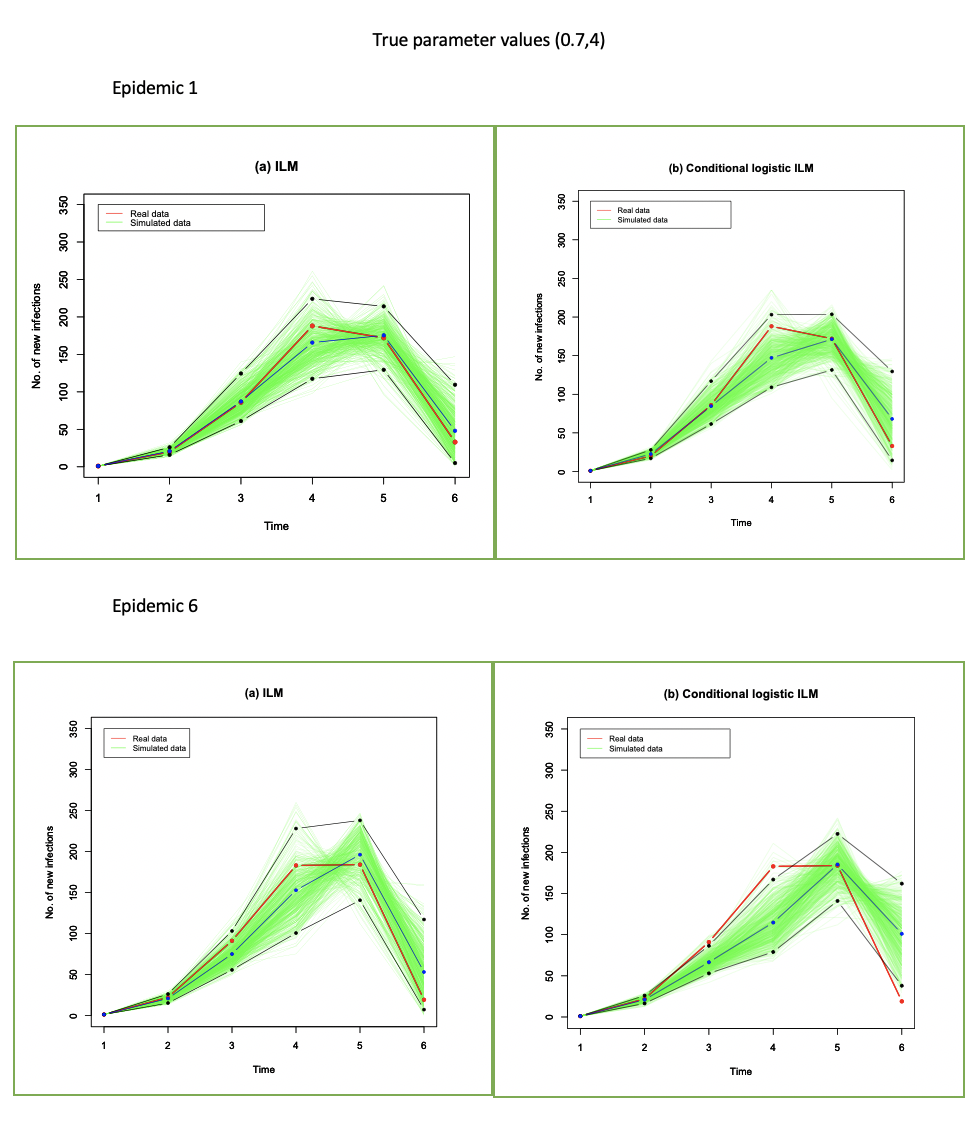}
		\caption{The posterior predictive distribution and 95\% credible intervals (black lines) for ILM and CL-ILM for various scenarios. The red line indicates the observed epidemic curve, the green lines are the 500 samples and the blue line is the mean of the sample.}
		\end{figure}

\begin{figure}[H]
		\centering
		\includegraphics[width=1\textwidth]{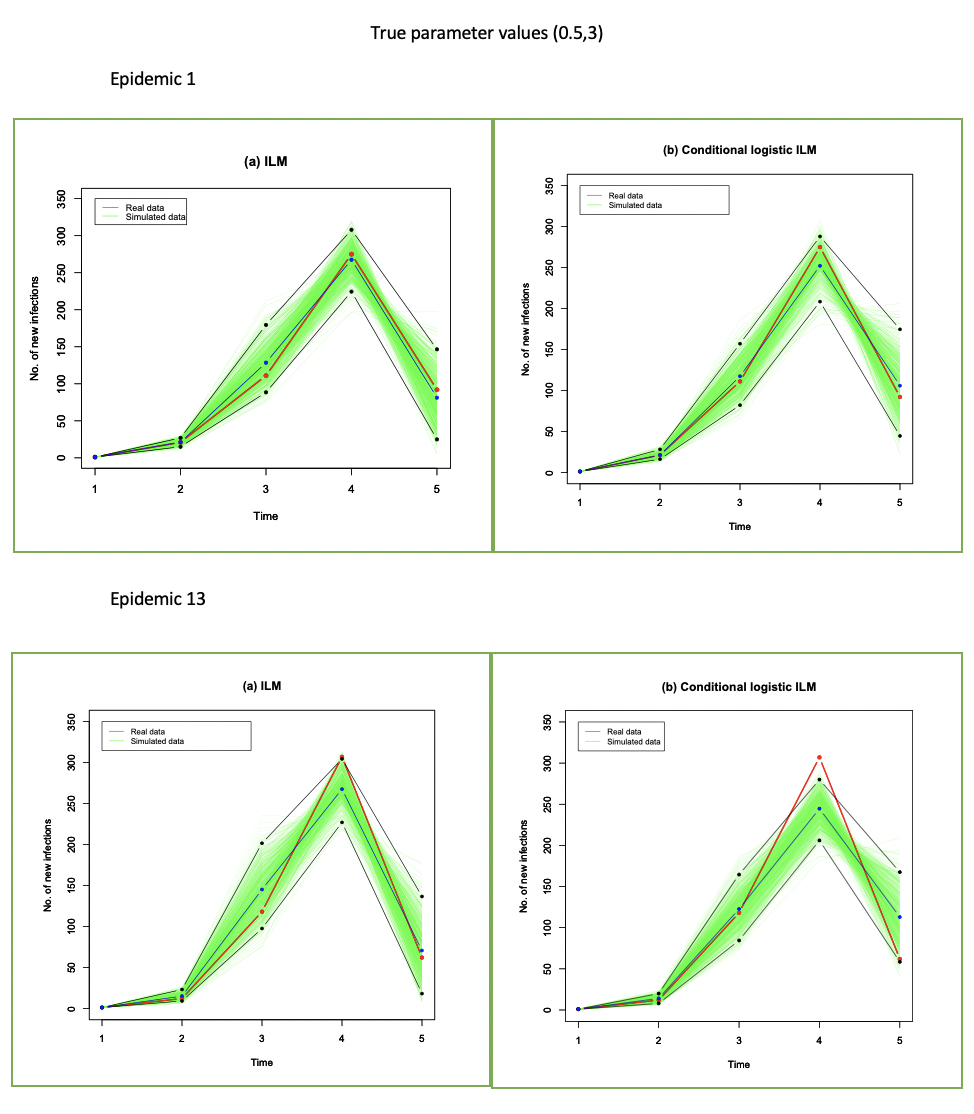}
		\caption{The posterior predictive distribution and 95\% credible intervals (black lines) for ILM and CL-ILM for various scenarios. The red line indicates the observed epidemic curve, the green lines are the 500 samples and the blue line is the mean of the sample. (continue) }
		\end{figure}

\begin{figure}[H]
		\centering
		\includegraphics[width=1\textwidth]{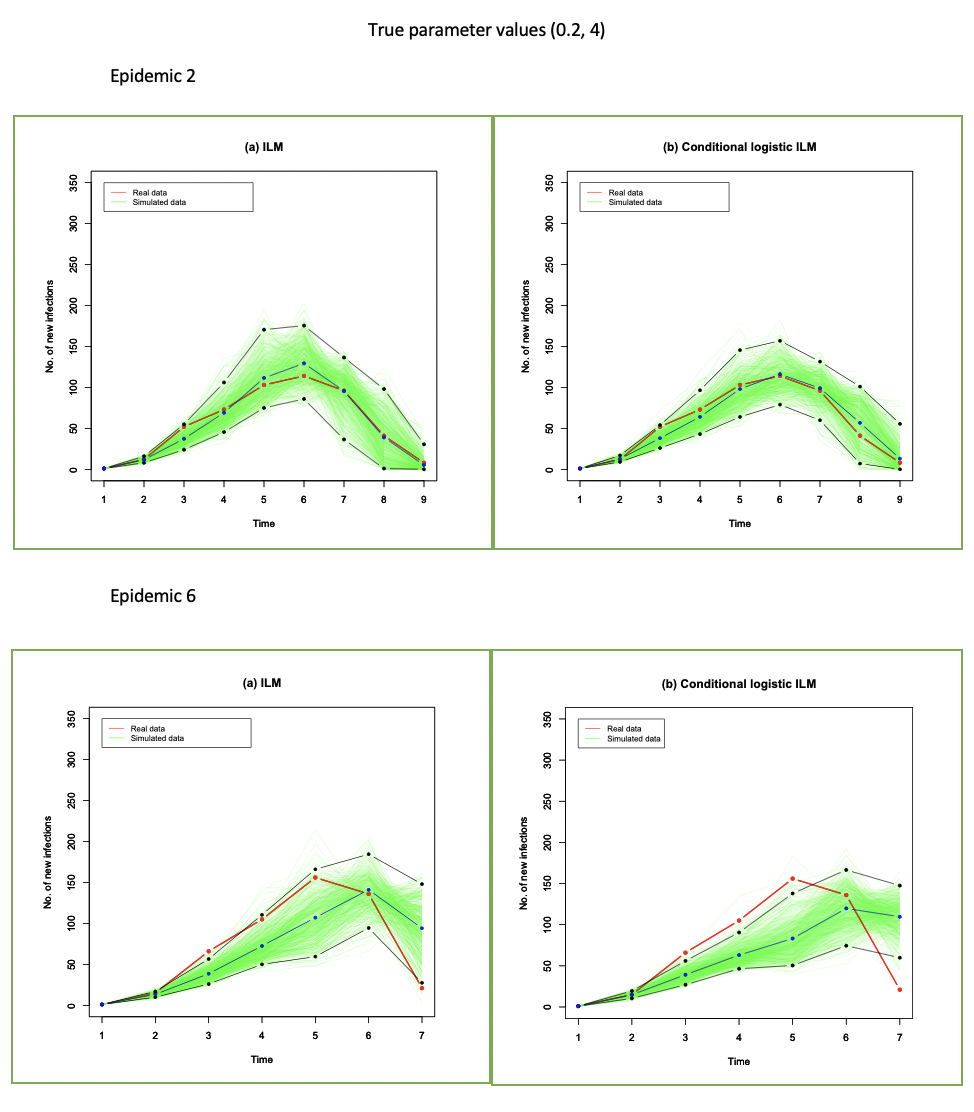}
		\caption{The posterior predictive distribution and 95\% credible intervals (black lines) for ILM and CL-ILM for various scenarios. The red line indicates the observed epidemic curve, the green lines are the 500 samples and the blue line is the mean of the sample.(continue) }
		\end{figure}

  \begin{figure}[H]
		\centering
		\includegraphics[width=1\textwidth]{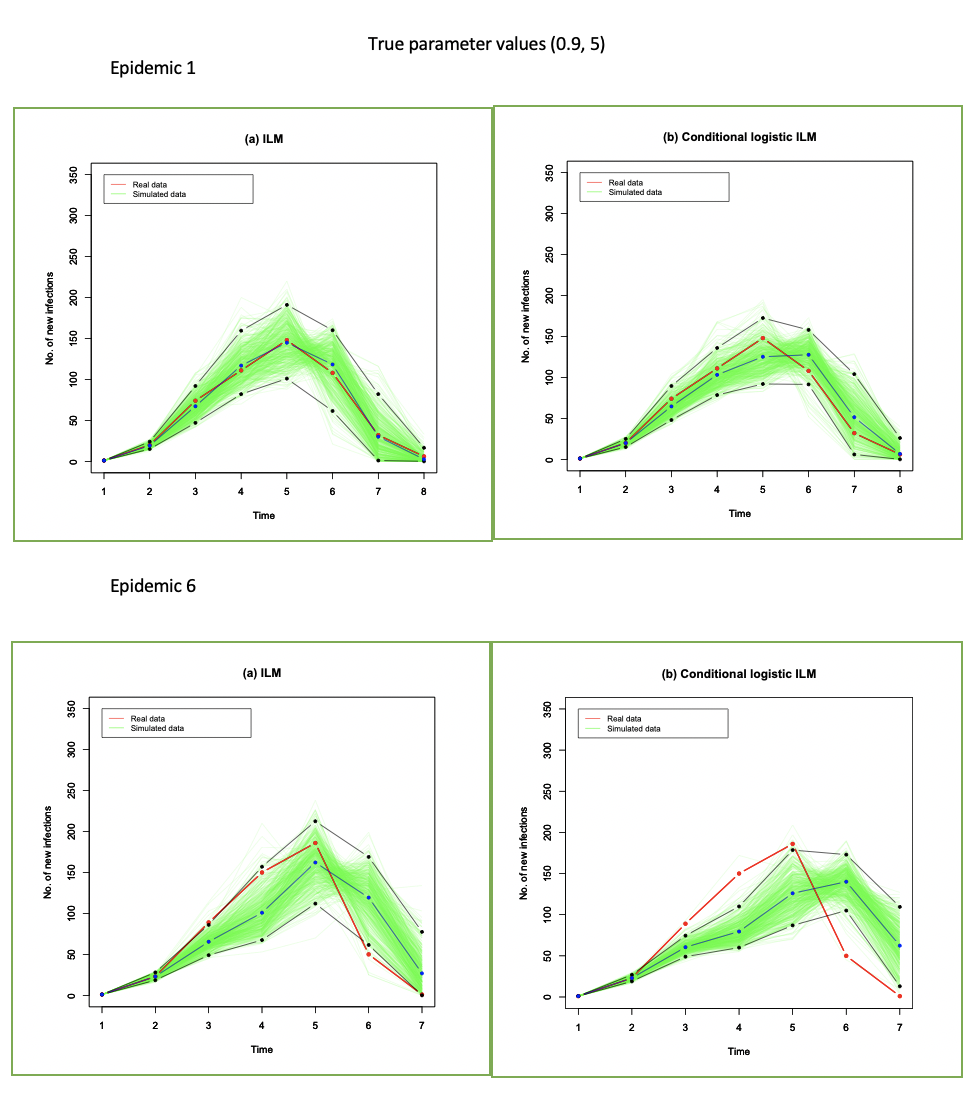}
		\caption{The posterior predictive distribution and 95\% credible intervals (black lines) for ILM and CL-ILM for various scenarios. The red line indicates the observed epidemic curve, the green lines are the 500 samples and the blue line is the mean of the sample. (continue) }
		\end{figure}

\subsection*{SIR}
  \begin{figure}[H]
		\centering
		\includegraphics[width=1\textwidth]{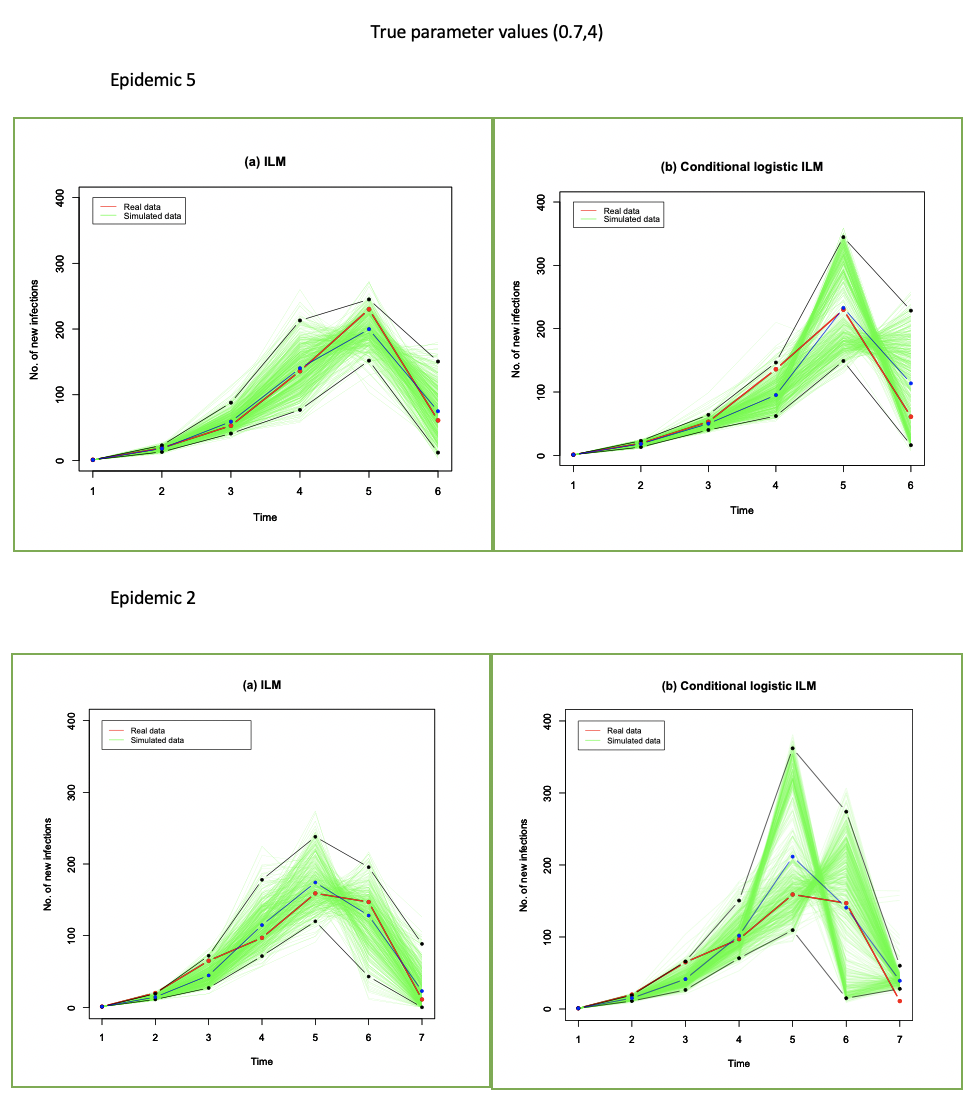}
		\caption{The posterior predictive distribution and 95\% credible intervals (black lines) for ILM and CL-ILM for various scenarios. The red line indicates the observed epidemic curve, the green lines are the 500 samples and the blue line is the mean of the sample.}
		\end{figure}

\begin{figure}[H]
		\centering
		\includegraphics[width=1\textwidth]{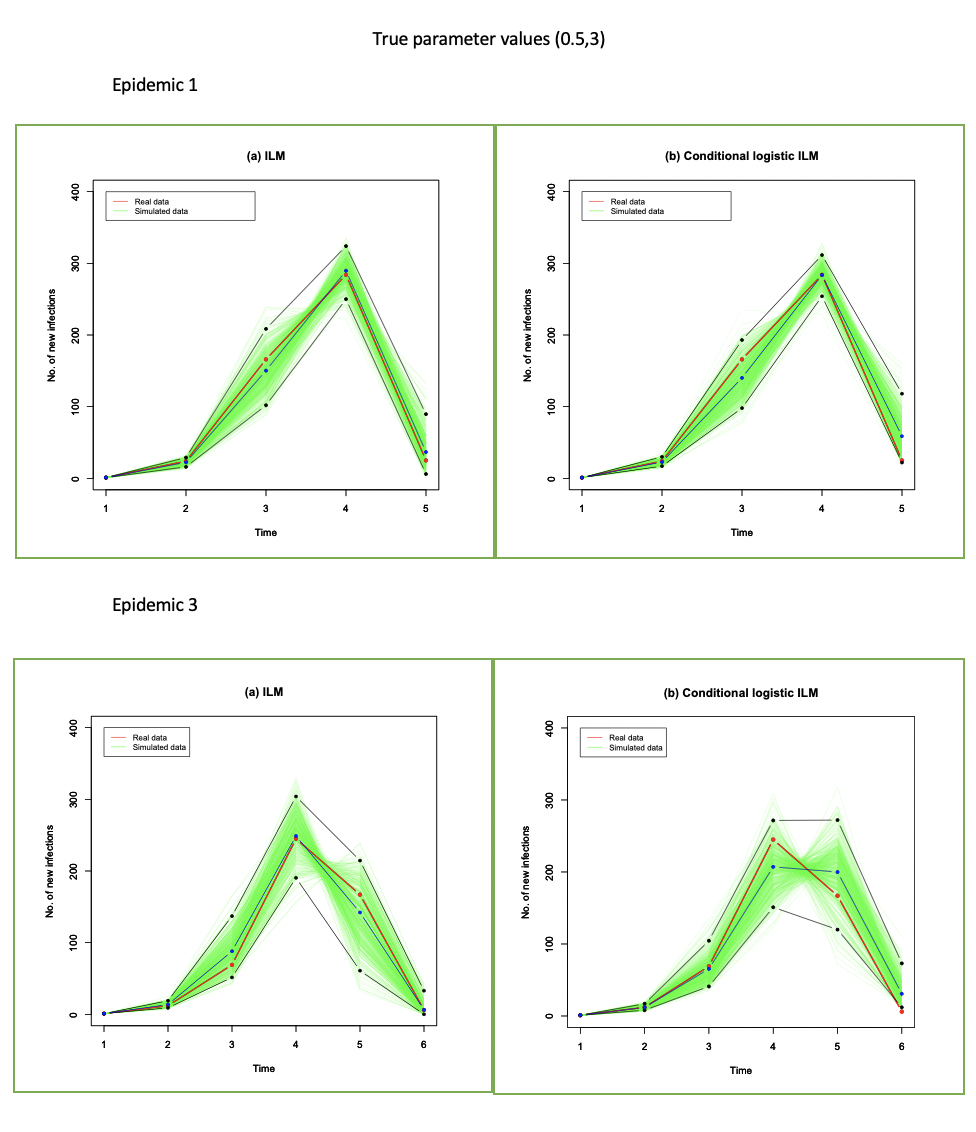}
		\caption{The posterior predictive distribution and 95\% credible intervals (black lines) for ILM and CL-ILM for various scenarios. The red line indicates the observed epidemic curve, the green lines are the 500 samples and the blue line is the mean of the sample. (continue) }
		\end{figure}

\begin{figure}[H]
		\centering
		\includegraphics[width=1\textwidth]{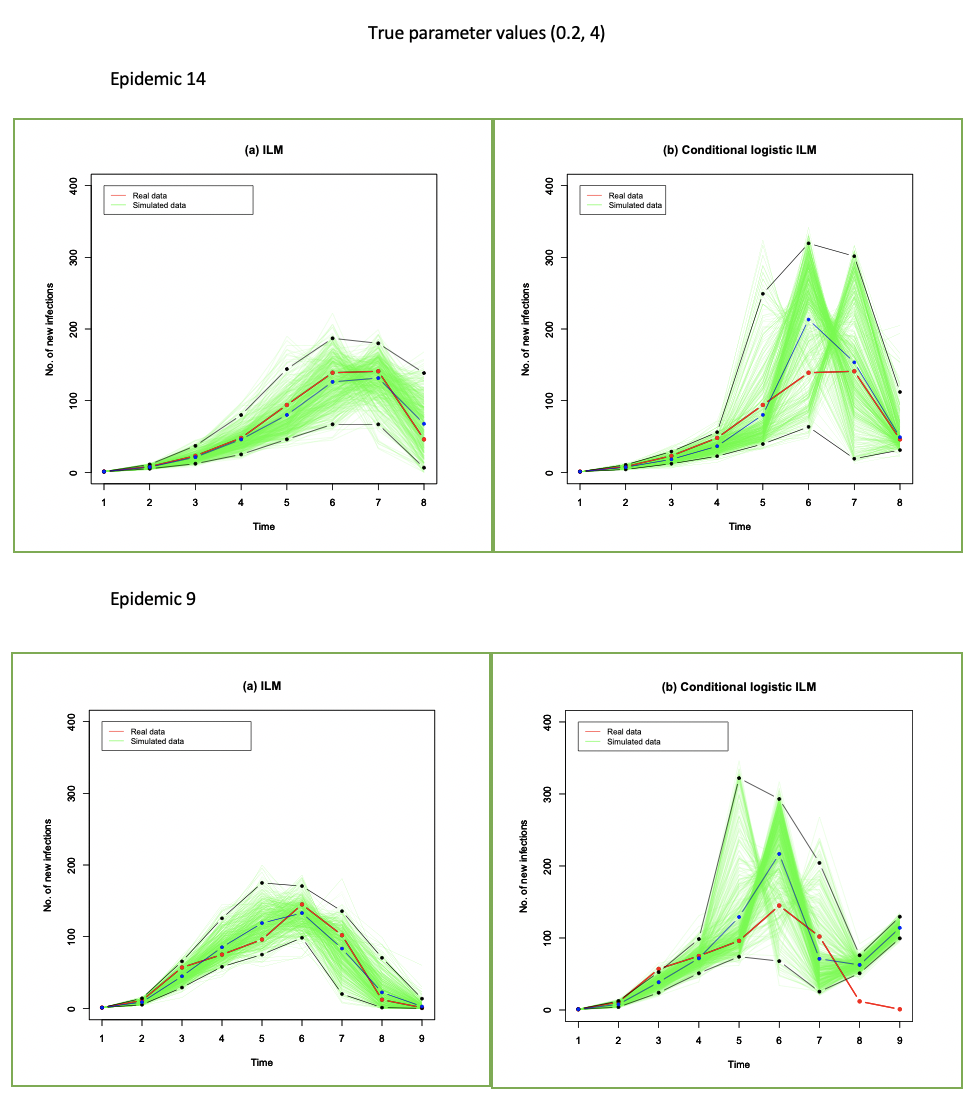}
		\caption{The posterior predictive distribution and 95\% credible intervals (black lines) for ILM and CL-ILM for various scenarios. The red line indicates the observed epidemic curve, the green lines are the 500 samples and the blue line is the mean of the sample.(continue) }
		\end{figure}

  \begin{figure}[H]
		\centering
		\includegraphics[width=1\textwidth]{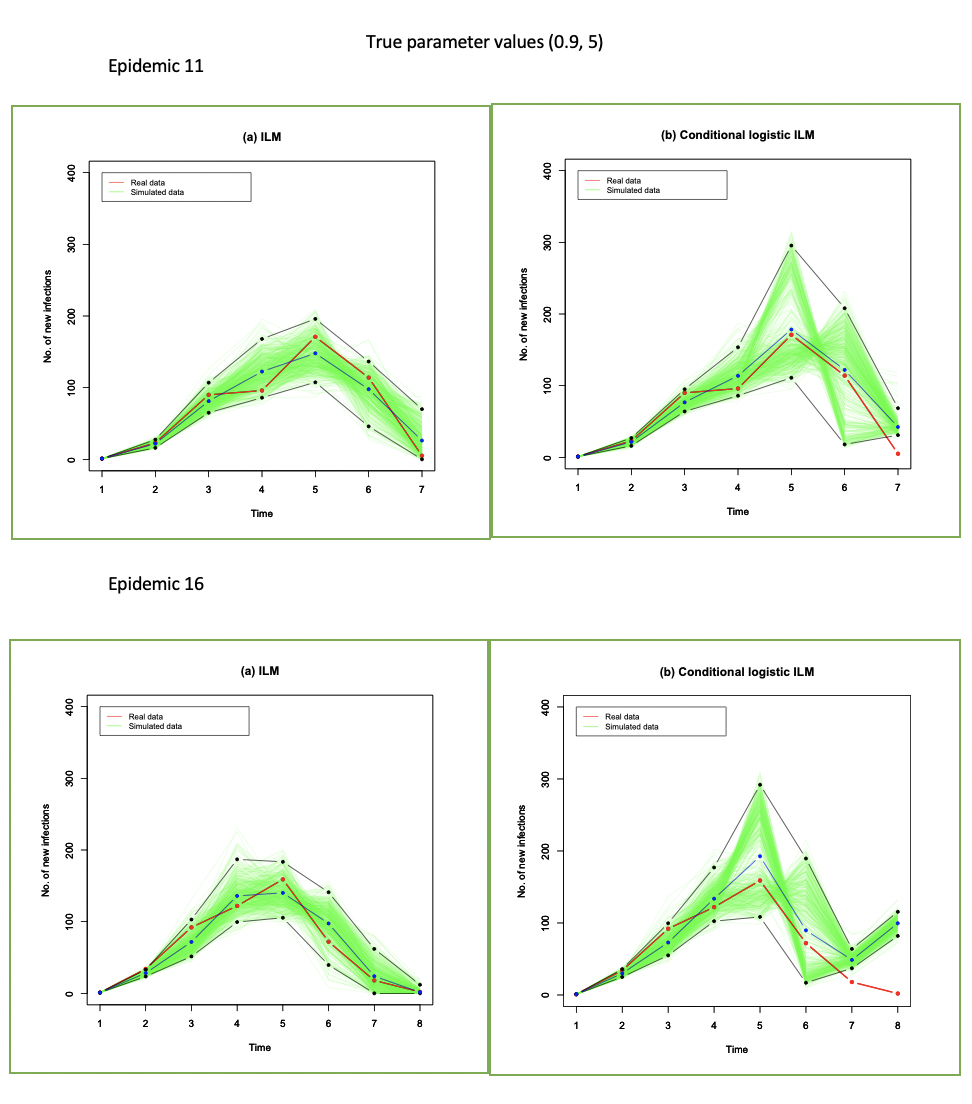}
		\caption{The posterior predictive distribution and 95\% credible intervals (black lines) of ILM and CL-ILM for various scenarios. The red line indicates the observed epidemic curve, the green lines are the 500 samples and the blue line is the mean of the sample. (continue) }
		\end{figure}

\subsection*{Semi-real Data}
\begin{figure}[H]
		\centering
		\includegraphics[width=1\textwidth]{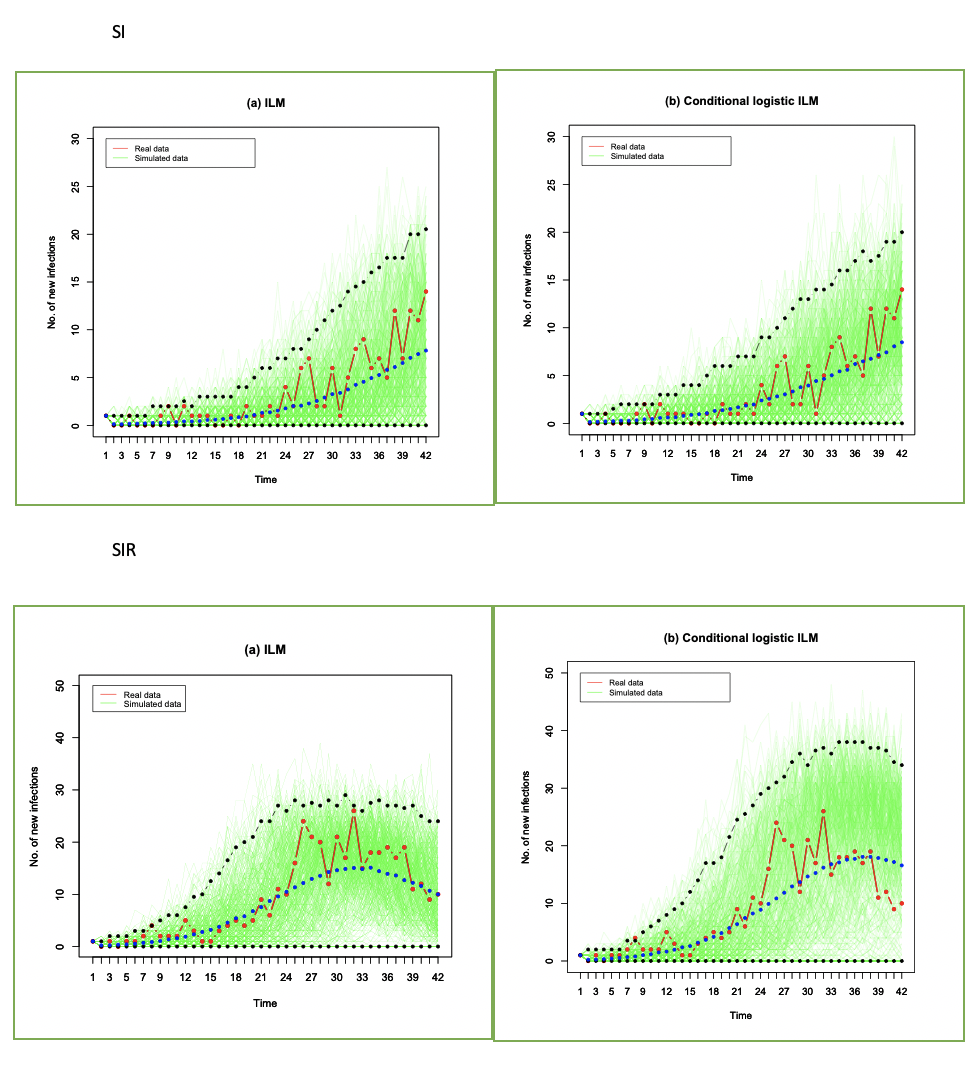}
		\caption{The posterior predictive distribution and 95\% credible intervals (black lines) of ILM and CL-ILM for semi-real data. The red line indicates the observed epidemic curve, the green lines are the 500 samples and the blue line is the mean of the sample. }
		\end{figure}

\end{document}